\documentclass[]{spie}  

\usepackage{amsmath,amsfonts,amssymb}
\usepackage{graphicx}
\usepackage{indentfirst}
\usepackage[colorlinks=true, allcolors=blue]{hyperref}

\title{Compressing radio interferometric visibility data into a probabilistic model using sparse Gaussian processes
}

\author[a, b]{Takafumi Tsukui}
\affil[a]{Kavli Institute for the Physics and Mathematics of the Universe (WPI), The University of Tokyo, Kashiwa, Chiba 277-8583, Japan}
\affil[b]{Center for Data-Driven Discovery, Kavli IPMU (WPI), UTIAS, The University of Tokyo, Kashiwa, Chiba 277-8583, Japan}

\authorinfo{Takafumi Tsukui\\T.T.: E-mail: tsukuitk23@gmail.com, Telephone: +81 8048397210\\}

\begin{document} 
\maketitle

\begin{abstract}
Next-generation radio interferometers will produce massive data volumes, making it impractical to store original visibility measurements and later combine observations in $uv$ spatial frequency space. Visibility measurements at similar $uv$ locations measure the same signal but different noise realizations. In principle, these measurements can therefore be compressed by storing only the inferred mean visibility and its uncertainty. We propose modeling the visibility with a sparse Gaussian process (GP) and storing the resulting compact probabilistic model rather than raw visibilities. Using simulated Atacama Large Millimeter/submillimeter Array (ALMA) observations, we demonstrate that the sparse GP is flexible enough to represent the visibilities and recover images with high fidelity. We estimate compression factors of $10^3-10^5$ for an 8-hour Square Kilometre Array (SKA)-Mid observation, with further gains expected by extending the GP input space to include the spectral axis. Beyond data compression, the model exploits correlations in $uv$ space, boosting the signal-to-noise ratio compared with independent grid averaging. Once trained, the model can predict visibility and its uncertainty at any desired $uv$ coordinates, allowing imaging with arbitrary fields of view and image resolutions. The model may also be incrementally updated with new observations while filtering outliers based on the prediction. 
\end{abstract}


\section{INTRODUCTION}
\label{sec:intro}  
An interferometer measures the visibility, the Fourier transform of the sky brightness, at spatial-frequency coordinates $uv$ set by each antenna pair baseline. With $N_A$ antennas, it measures the visibilities from $N_B=N_A(N_A-1)/2$ baselines every few seconds for each frequency channel and polarization. Earth rotation sweeps these samples across $uv$ space, increasing the Fourier sampling needed to reconstruct the sky brightness distribution. The number of baselines $N_B$ becomes large for next-generation interferometers such as SKA-Low ($N_A=512$), SKA-Mid ($N_A=197$)~\cite{dewdneySKA12022}, ngVLA ($N_A=263$)~\cite{selinaScience2018}, ALMA 2040 ($N_A=200–300$)~\cite{Summary}. This results in data rates of hundreds of GB/s~\cite{selinaSystem2019}, making it increasingly challenging to store the calibrated visibilities.

The resulting visibility set is enormous in storage volume but highly redundant. Many measurements sample the visibility at nearly the same region of $uv$ space, with different thermal-noise realizations. As a result, most stored bits describe measurement noise rather than information about the sky distribution. The information most relevant for imaging is in the mean visibility and its uncertainty, not in every individual visibility sample. Storing all visibilities is therefore not an efficient representation of the information essential for interferometric imaging.

For SKA, the current plan is to retain only imaging products and gridded visibilities and to delete the calibrated visibilities after a fixed archive period~\cite{SKAO}. Gridding greatly reduces the storage volume by replacing many noisy visibility measurements with a mean visibility and uncertainty on a fixed $uv$ lattice~\cite{Golap2016MSUVBIN}. This is sufficient for analyzing a single observation, but it limits the value of the data for future studies. Many science cases require combining observations to reach greater depth or spatial resolution. Once the visibilities have been averaged onto a fixed $uv$ grid, the data become tied to that grid, which sets the field of view and image resolution. Information below the grid scale is lost and cannot be recovered. As a result, combining datasets becomes less flexible, and the propagation of uncertainties becomes more difficult. A similar limitation applies to baseline-dependent averaging, which fixes the maximum $uv$ distance over which each baseline is averaged, and hence the achievable field of view~\cite{wijnholdsBaselinedependent2018}.

A natural alternative is to model the visibility as a continuous function of $uv$ coordinates. Gaussian processes (GPs) are well suited to this task. They provide a flexible, non-parametric model of the visibility together with principled uncertainty estimates~\cite{Rasmussen2006GPML}. The model exploits correlations between neighboring visibilities in $uv$ space to infer the underlying visibility, improving the signal-to-noise ratio relative to simple grid averaging. Once trained, the model provides a continuous estimate of the visibility together with its posterior uncertainty, conditioned on all observed data, at any $uv$ coordinate within the sampled range. The model is not tied to a fixed grid: users can choose the imaging cell size and field of view at use time, without re-fitting or accessing the original visibilities.

The model must remain compact and trainable on datasets containing billions of visibilities. Exact GPs do not meet the requirement: their memory, computational cost, and model size all scale as $N_{\rm data}^2$, $N_{\rm data}^3$, $N_{\rm data}$ respectively. Sparse Gaussian processes address this problem by replacing the full dataset with $M \ll N_{\rm data}$ inducing points. The model size scales as $\mathcal{O}(M^2)$, while converging to the exact GP as $M$ increases~\cite{titsiasVariational2009}. A practical method must also train without loading the full dataset into memory. Conventional sparse GP methods still require access to all $N_{\rm data}$ visibilities when optimizing the model. Sparse variational Gaussian processes (SVGPs) overcome this limitation by training from small data batches, allowing the data to be streamed during training while keeping the memory cost independent of $N_{\rm data}$~\cite{hensmanGaussian2013, hensmanScalable2014}.

In this paper, we evaluate sparse variational Gaussian processes as a compact representation of interferometric visibility data using a mock Atacama Large Millimeter/submillimeter Array (ALMA) observation (Sec.~\ref{subsec:subsec1}). We describe the sparse variational Gaussian process framework and its stochastic training procedure, which allows visibility datasets much larger than available memory to be processed efficiently (Sec.~\ref{sec:svgp_model} and \ref{sub:implementation}). In Sec.~\ref{sec:results}, we assess the reconstructed visibility, the quality of images generated from the compressed representation, and the compression factor. We also compare the method with existing visibility-compression approaches and discuss directions for future work.

\section{Method}
\subsection{Mock data}\label{subsec:subsec1}
We generated mock ALMA visibility data with the CASA task \textsc{simalma}, using the H$\alpha$ image of M51 as the input sky model\footnote{\url{https://casaguides.nrao.edu/images/3/3f/M51ha.fits.txt}}. We adopted the ALMA configuration file \texttt{alma.cycle8.3.cfg}, a central frequency of 330.076~GHz, a bandwidth of 50 MHz, a precipitable water vapour of 0.6~mm, and a total on-source integration time of 1800~s. We assigned the input image an angular cell size of 0.02~arcsec. To test different signal-to-noise (S/N) regimes, we generated three simulation sets by scaling the input sky brightness by factors of 1, 1/10, and 1/30, corresponding to high-S/N, low-S/N, and noise-dominated cases.

\subsection{Sparse Gaussian process model}
\label{sec:svgp_model}

We use a sparse Gaussian-process model for the real and imaginary parts of the visibility. We assign independent Gaussian-process priors to the two components. For either component, denoted here by $f$, we model
\begin{equation}
f(\mathbf{x}) \sim \mathcal{GP}\!\left(0,\, k(\mathbf{x}, \mathbf{x}')\right),
\end{equation}
where the mean is zero and the covariance function $k$ sets how strongly the visibility values at two $uv$-plane locations, $\mathbf{x}=(u,v)$ and $\mathbf{x}'=(u',v')$, are correlated. For any finite set of $uv$ coordinates $\mathbf{X}=\{\mathbf{x}_i\}_{i=1}^n$, the corresponding function values $\mathbf{f}$ are jointly Gaussian,
\begin{equation}
\mathbf{f} \sim \mathcal{N}(\mathbf{0}, \mathbf{K}),
\qquad
K_{ij}=k(\mathbf{x}_i,\mathbf{x}_j).
\end{equation}
Thus, the kernel determines which visibility functions are more probable under the prior.

We use a Mat\'ern-5/2 kernel with separate lengthscales along the $u$ and $v$ directions,
\begin{equation}
k(\mathbf{x}, \mathbf{x}')
=
\sigma_f^2
\left(1 + \sqrt{5}\,r + \frac{5}{3}r^2\right)
\exp(-\sqrt{5}\,r),
\qquad
r^2
=
\frac{(u-u')^2}{\ell_u^2}
+
\frac{(v-v')^2}{\ell_v^2},
\end{equation}
where the parameter $\sigma_f^2$ sets the prior variance, while $\ell_u$ and $\ell_v$ set the correlation lengths along the two $uv$-plane directions. The Matérn-5/2 kernel imposes a moderately smooth prior in which the visibility varies smoothly across the $uv$ plane, with sample functions that are twice mean-square differentiable. This choice is consistent with the expectation that visibility varies continuously across the $uv$ plane for a finite-extent sky, while remaining flexible enough to accommodate structure supported by the data.

An exact Gaussian process conditions on all $N_{\rm data}$ visibilities, which is too expensive for large datasets and leads to no reduction of the data. The sparse Gaussian process therefore approximates it with $M \ll N_{\rm data}$ inducing points which act as a compact summary of the visibility. The inducing-point locations in the $uv$ plane are denoted by $\mathbf{Z}$, and the corresponding visibility values $\mathbf{u}$ are described by the variational posterior distribution
\begin{equation}
q(\mathbf{u}) = \mathcal{N}(\mathbf{m}, \mathbf{S}),
\end{equation}
where $\mathbf{m}$ is the mean vector and $\mathbf{S}$ is the covariance matrix. The model combines this variational distribution with the kernel covariance to predict the visibility at any $uv$ coordinate.

The training objective is the evidence lower bound (ELBO) computed for $N_{\rm data}$ data points ($\mathbf{x}_n$, $y_n$)~\cite{hensmanGaussian2013, hensmanScalable2014}:
\begin{equation}
\mathcal{L} =
  -\frac{N}{2}\log(2\pi\sigma^2)
  - \frac{1}{2\sigma^2}\sum_{n=1}^{N}\!\left[\bigl(y_n - \mathbb{E}_q(f_n)\bigr)^2 + \mathrm{Var}_q(f_n)\right]
  - \mathrm{KL}\!\left[q(\mathbf{u}) \,\|\, p(\mathbf{u})\right].
\end{equation}
where $\mathbb{E}_q(f_n)$ and $\mathrm{Var}_q(f_n)$ are the predictive mean and variance of the GP at $\mathbf{x}_n$ under $q(\mathbf{u})$, and $p(\mathbf{u}) = \mathcal{N}(\mathbf{0}, \mathbf{K}_{ZZ})$ is the GP prior evaluated at the inducing locations. The first two terms are the standard Gaussian log likelihood and measure how well the model reproduces the measured visibilities. The KL term penalizes deviations of $q(\mathbf{u})$ from the GP prior at the inducing locations, which keeps the inferred visibility consistent with the assumed kernel structure. We maximise $\mathcal{L}$ jointly over the inducing locations $\mathbf{Z}$, the variational parameters $(\mathbf{m},\mathbf{S})$, the kernel hyperparameters $(\sigma_f^2,\ell_u,\ell_v)$, and the noise variance $\sigma^2$, which together fully describe the trained model and total $M(M+1)/2 + 3M + 4 = \mathcal{O}(M^2)$ parameters, independent of $N_{\rm data}$.

\subsection{Implementation}\label{sub:implementation}
We implement the model with GPflow~\cite{GPflow2017}, whose \texttt{SVGP} class provides the sparse variational construction described in Section~\ref{sec:svgp_model} and the ELBO objective, which we maximise by stochastic variational inference~\cite{hensmanGaussian2013, hensmanScalable2014}. Below we summarize the practical points specific to modeling visibility data at scale. \\
\textbf{Exploiting the Hermitian symmetry and data normalization:}
Because the sky brightness is real-valued, the visibility has Hermitian symmetry,
\begin{equation}
V(-u,-v) = V^{*}(u,v).
\end{equation}
We fold the data onto the $u \geq 0$ half-plane and model the field there, halving the input space without information loss. We model the real and imaginary parts independently. Each component is normalized by its own robust scale, defined as the difference between its 99th and 1st percentiles, so that the targets are of order unity. The scale factors are stored with the model and applied when converting predictions back to physical units.\\
\textbf{Minibatching with shuffled data:}
The data-fit term in the ELBO is a sum over the $N$ visibilities, so a uniformly drawn minibatch of size $b$ gives an unbiased estimate after rescaling by $N_{\rm data}/b$~\cite{hensmanGaussian2013}. We therefore train the model by stochastic variational inference, streaming the data through the optimizer in minibatches of $b$, so the full visibility data $N_{\rm data}$ is never held in memory at once. The total number of iterations $n_{\rm iters}$ is set by the average number of visits per data point $n_{\rm visit}$, $n_{\rm iters}=(N_{\rm data}/b) n_{\rm visit}$. We pre-shuffle the visibility array to ensure each minibatch is a uniform sample of $uv$ space, since Measurement Sets are otherwise stored in time--baseline order.\\
\textbf{Initial inducing points and optimization:}
Inducing locations are initialized by uniform sampling in polar coordinates $(r, \theta)$ on the $u \geq 0$ half-plane (radial density $\propto 1/r$ in the $(u,v)$ plane), which concentrates points near the $uv$ origin to match the natural concentration of visibility measurements. The variational mean $\mathbf{m}$ is initialized to zero and the variational covariance $\mathbf{S}$ to the identity. We optimize the variational parameters $(\mathbf{m}, \mathbf{S})$ using natural gradients~\cite{salimbeniNatural2018} while the inducing locations $\mathbf{Z}$, kernel hyperparameters, and noise variance are optimized with Adam~\cite{kingmaAdam2017} using a learning rate of $10^{-2}$. At each iteration, the natural-gradient and Adam updates are applied sequentially to the same minibatch. 

\textbf{Number of inducing points, iterations, and batch size}: In this study, we used $M = 500$ and $b = 4096$, which fit comfortably within the CPU and memory limits of a typical laptop. Larger $M$ is better for more complex imaging, particularly wide-field, which requires higher-resolution recovery in $uv$ space. Larger values of $b$ improve the accuracy of the unbiased likelihood estimate. We found that $n_\mathrm{visit}=100$ is long enough to reach convergence for all S/N setups.

\section{Results and discussion}
\label{sec:results}
\begin{figure}
    \centering
    \includegraphics[width=1\linewidth]{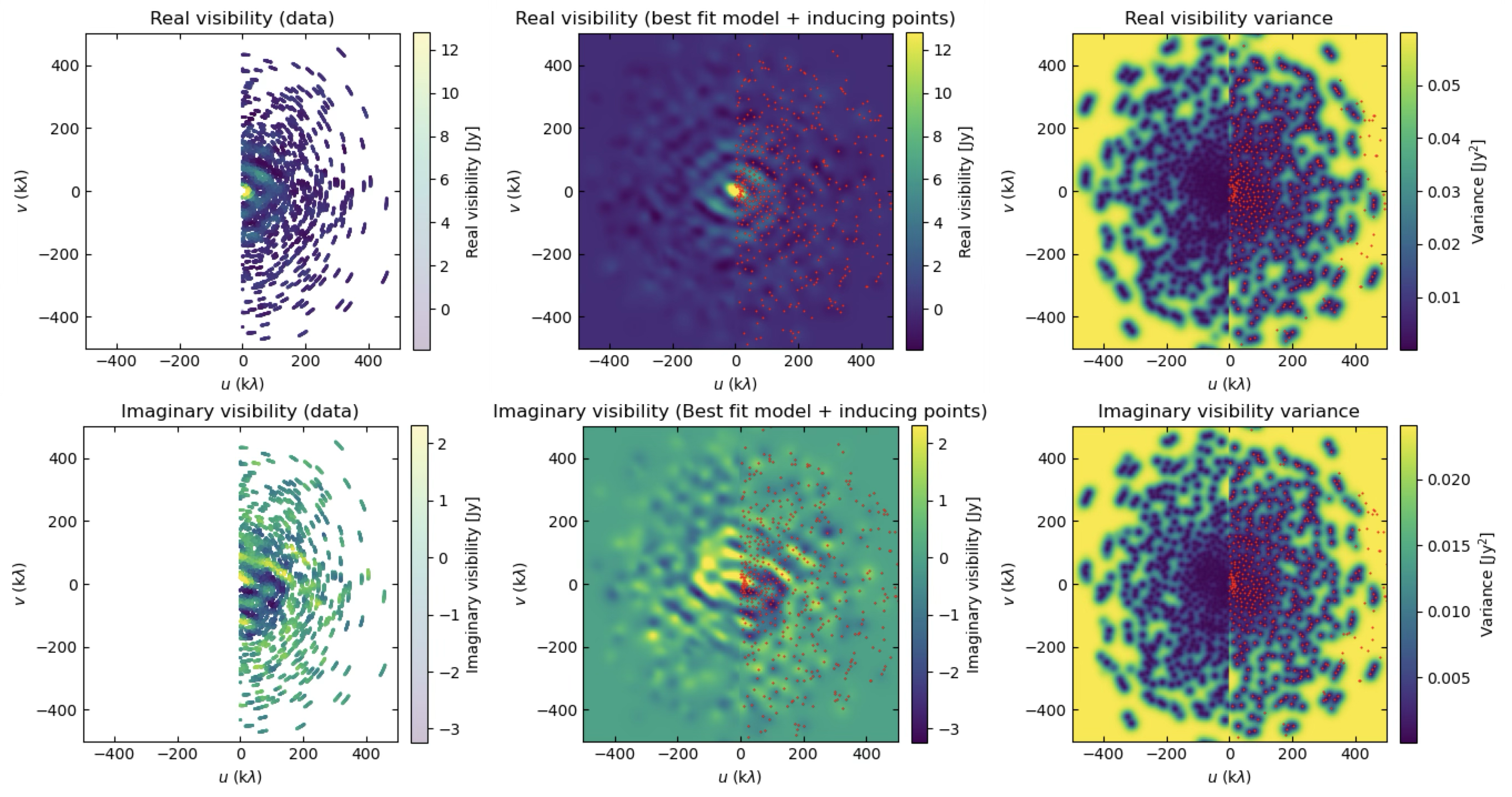}
    \caption{Visibility data (left), recovered mean (middle), and posterior variance (right) for the real (top) and imaginary (bottom) components of the visibility. Red points indicate the locations of the best-fit inducing points.}
    \label{fig:image0}
\end{figure}

\begin{figure}
    \centering
    \includegraphics[width=1\linewidth]{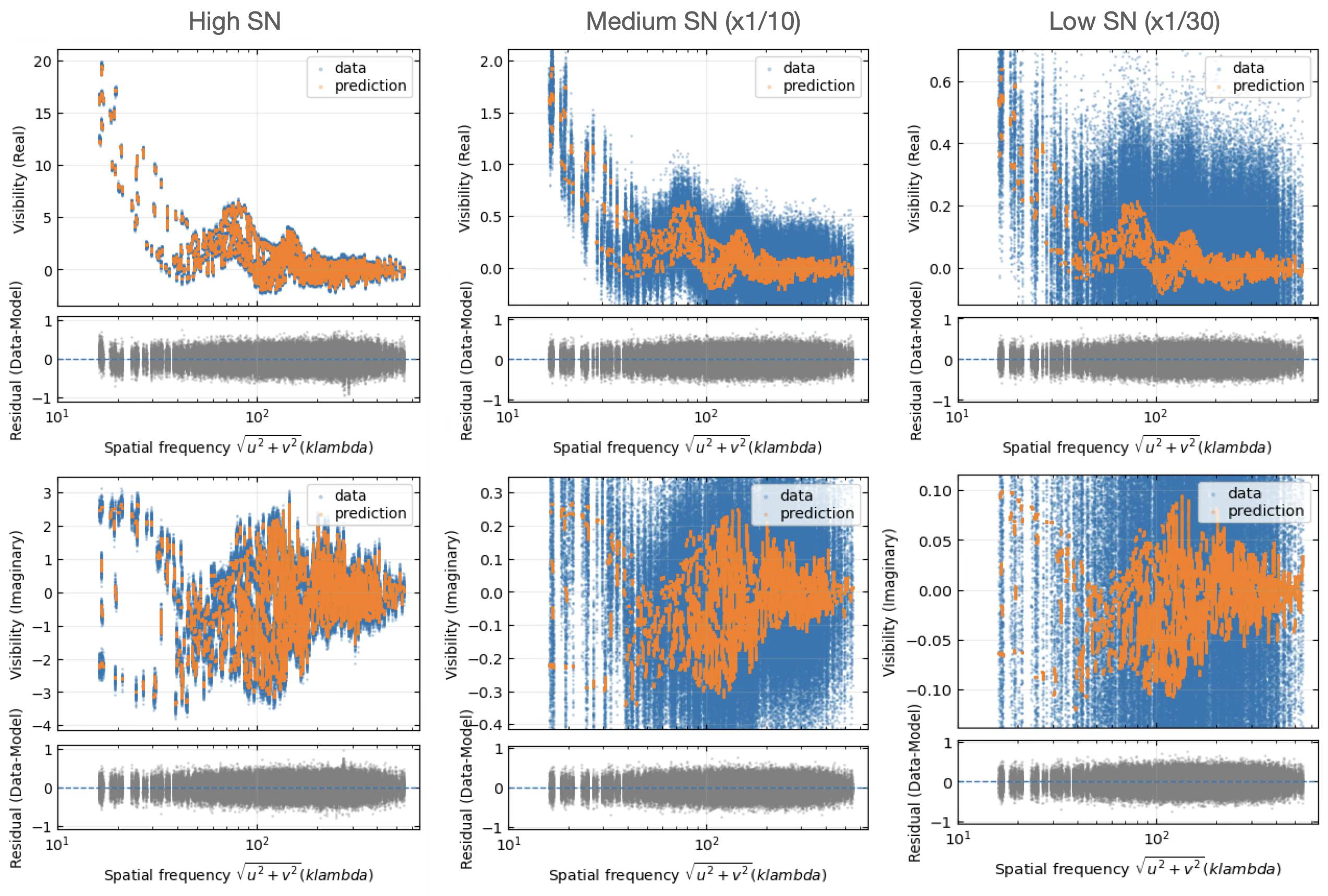}
    \caption{Noisy visibility data used for modeling (blue) and the recovered mean are plotted against spatial frequency. The columns correspond to high (left), medium (middle), and low (right) S/N cases.}
    \label{fig:image1}
\end{figure}

\textbf{Mean visibility recovery}: Fig.~\ref{fig:image0} shows the measured visibility data, the recovered mean visibility, the posterior variance, and the inducing-point locations for the high signal-to-noise case. The complex visibility structure is accurately described by only 500 inducing points. Several inducing points appear to contribute little to the reconstruction, suggesting that $M=500$ is a conservative choice and that comparable fidelity is achievable with even fewer inducing points. This is further demonstrated in Fig.~\ref{fig:image1}, which compares the recovered mean visibility with the original visibility measurements for all signal-to-noise ratios. Even at low S/N, the model successfully recovers the mean visibility. 

\begin{figure}
    \centering
    \includegraphics[width=1\linewidth]{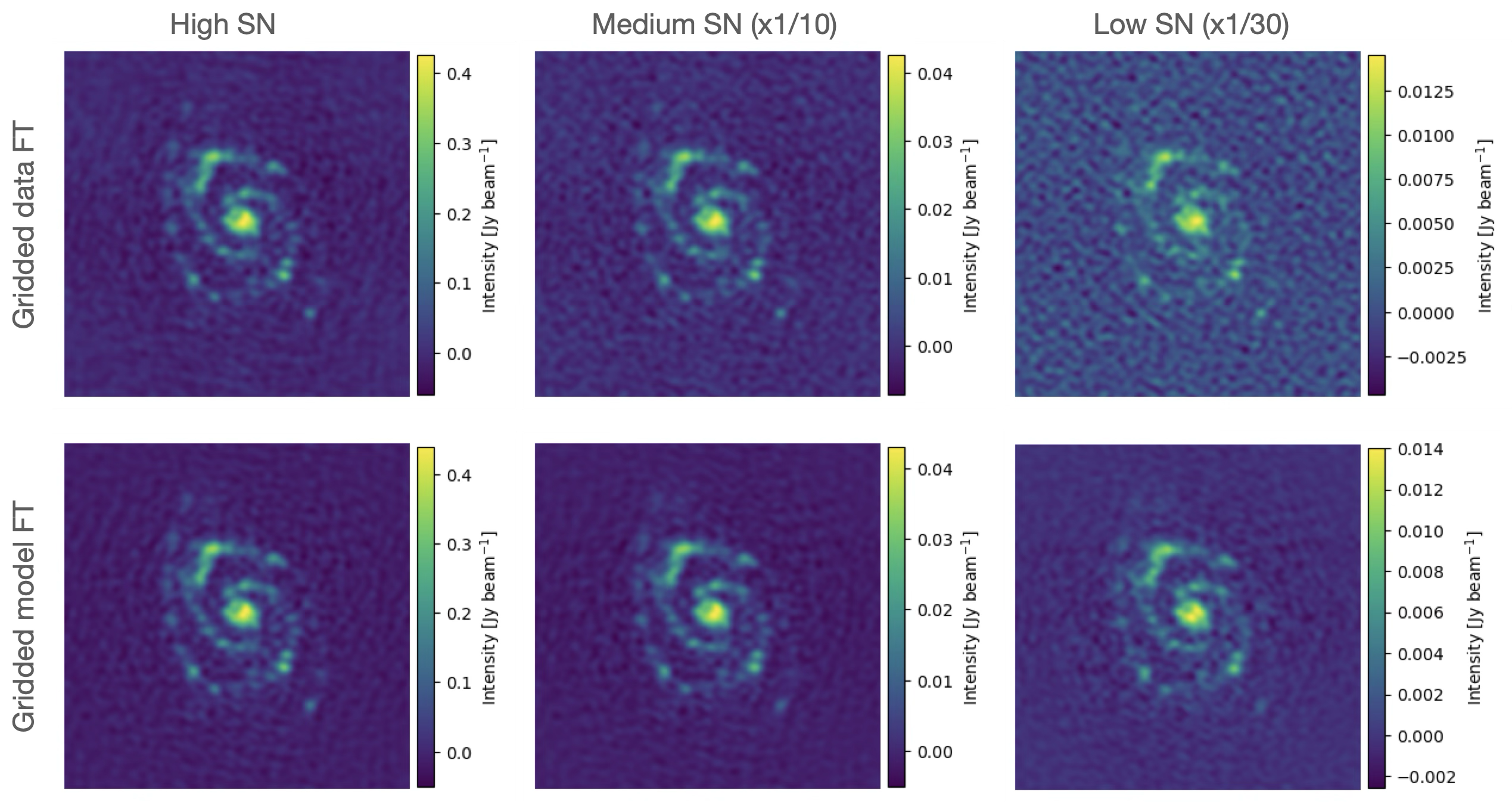}
    \caption{
    Top: Uniformly weighted image obtained by gridding the original visibilities, averaging them within each $uv$-grid cell, and applying a discrete Fourier transform.
    Bottom: Uniformly weighted image obtained from the model visibilities evaluated on the same $uv$-grid cells and transformed using the same procedure.
    }
    \label{fig:image2}
\end{figure}

\begin{figure}
    \centering
    \includegraphics[width=1\linewidth]{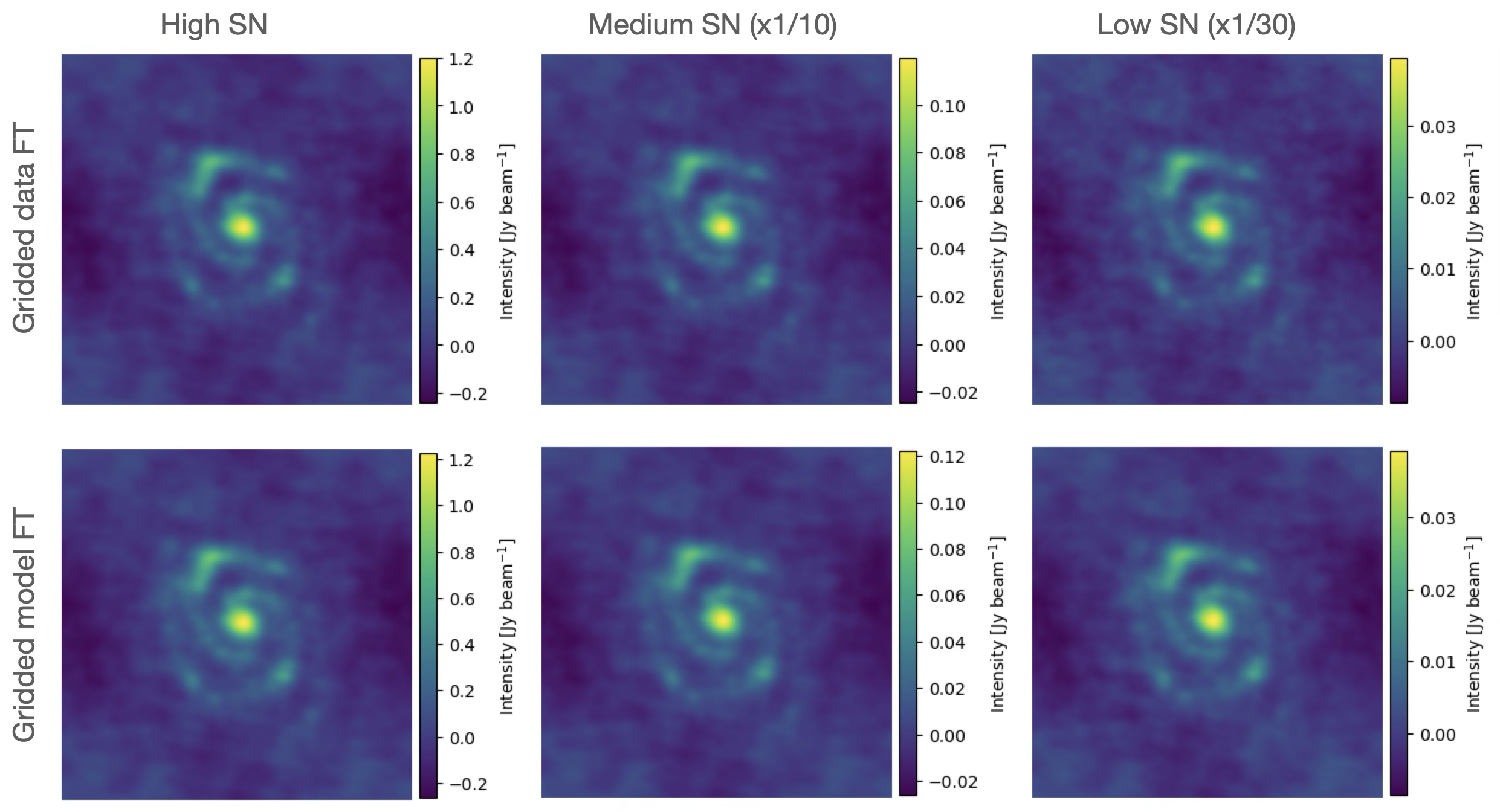}
    \caption{
    Same as Figure~\ref{fig:image2}, but using natural weighting.
    }
    \label{fig:image3}
\end{figure}

\textbf{Image recovery}: To assess image quality, we do not consider the full imaging processes, including gridding, weighting, and deconvolution, as these are separable problems that have been studied extensively~\cite{hogbomAperture1974, schwarzMathematicalstatistical1978, briggsHigh1995, yeOptimal2020}. Instead, we compare image quality in a controlled way. We image the original measurements by averaging the visibilities within each $uv$-grid cell and applying a discrete Fourier transform. We image the modelled visibility by evaluating the posterior mean at the same $uv$-grid cells and applying the same transform. For both cases, we did not deconvolve, and we consider two weighting schemes: uniform weighting, with $w = 1$ for grid cells containing data and $w=0$ otherwise, and natural weighting, with $w=N$, where $N$ is the number of measurements in each grid cell. The image reconstructed from the posterior mean has higher fidelity under uniform weighting, recovering the visible tidal tail in the low-S/N case that is not recovered from the grid-averaged data (Fig.~\ref{fig:image2}). In contrast, the naturally weighted images are nearly identical across all S/N cases (Fig.~\ref{fig:image3}). This result suggests that the data already provide sufficient S/N on the spatial scales emphasized by natural weighting. It also highlights the value of a continuous visibility model, allowing arbitrary $uv$-grids to be constructed after inference for different imaging resolutions and fields of view.

\textbf{Data compression}: For 500 inducing points, the trained sparse GP model occupies 1.0~MB per component (the real and imaginary parts modelled independently), or 2.0~MB in total. For the 1800~s M51 simulation with 10~s integration and 43 antennas ($N_{\rm data} = 162{,}540$), the calibrated measurement set (complex visibilities and $(u, v)$ coordinates) is 5.2~MB on disk (32 bytes per visibility). This corresponds to a compression factor of $\sim 2.5$. Because the stored model size scales as $\mathcal{O}(M^2)$ and is independent of $N$, the compression factor grows linearly with dataset size. For example, a typical 8-hour SKA-Mid observation with 197 dishes and 0.14~s time sampling would produce approximately 127~GB of visibility data, ignoring frequency and polarization axes~\cite{bonaldiSquare2018}. At the same $M = 500$, the model would compress this by a factor of $\sim 6\times10^4$. Increasing the number of inducing points increases the stored model size as $\mathcal{O}(M^2)$. For example, the combined real and imaginary model occupies $\sim8$~MB at $M=1000$ and $\sim32$~MB at $M=2000$. Even at these larger values of $M$, the compression factor for SKA-scale datasets remains in the range $10^3–10^5$.

\textbf{Future work}: We focus here on compression and reconstruction performance in the spatial-frequency ($u,v$) domain. The data are also oversampled along the spectral axis $\nu$. Extending the Gaussian-process input space to $\mathbf{x}=$($u$,$v$,$\nu$) would be straightforward and could exploit correlations between frequency channels, yielding substantially greater compression. A proof of concept for modeling spectral correlations with Gaussian processes has been demonstrated in the one-dimensional spectral domain for separating the 21-cm signal from foreground emission~\cite{mertensStatistical2018}.

We did not systematically explore inducing-point initialization. Preliminary tests indicate that the initialization has a substantial effect on training stability and reconstruction quality. We found that a simple radially uniform initialization in $(r, \theta)$ space produced stable optimization and accurate reconstructions, whereas uniform areal sampling in the $uv$ plane performed less well. We also tested the conditional-variance initialization method \cite{burt2020gpviconv} and found that it achieved comparable reconstruction fidelity with fewer inducing points.

In this work, we assumed a single noise variance shared by all visibilities. In practice, however, the noise can vary across the $uv$ plane, between baselines, and between observing epochs, and could instead be modelled with a heteroscedastic noise model. Such a model would also enable iterative refinement of the training process by progressively identifying and excluding measurements whose deviations from the predicted mean cannot be explained by the locally inferred thermal noise, affected by residual radio-frequency interference (RFI), calibration errors, or unusually noisy antennas.

The next step is application to realistic datasets from the target instrument. A full scaling study --- including hardware requirements, computational optimization for realistic data volumes, model complexity, and the trade-off between batch size $b$ and visit count $n_\mathrm{visit}$ --- is left to future work. The environmental and financial costs of long-term data storage must be weighed against the compute costs of model training and retraining, as well as the scientific value provided by the model. A full assessment of these trade-offs is beyond the scope of this work.

\textbf{Comparison with other compression methods}: Existing visibility-compression methods compress the stored visibility measurements. Their compression factors are therefore largely determined by the encoding scheme or by the effective complexity of the visibility data. Examples include lossy quantization (Dysco)~\cite{offringaCompression2016}, error-controlled compression (MGARD)~\cite{gongMGARD2023, dodsonOptimising2025}, baseline-dependent low-rank approximations~\cite{atemkengLossy2023} and lossless coding (Sisco)~\cite{offringaLossless2026}. These methods typically achieve compression factors of a few to a few tens while preserving the sampled visibilities. Another approach closest in spirit is compressive sensing, which compresses the antenna signals into random rank-one projections from which an image is reconstructed~\cite{leblancMROP2025}. Crucially, it compresses toward the image — its projections are tied to an imaging operator and weighting that are fixed once the data are compressed.

By contrast, our approach stores a probabilistic model of the visibility rather than the visibility samples themselves. The model scales as $\mathcal{O}(M^2)$ and is independent of $N_{\rm data}$, so the compression factor increases with dataset size. The gain is therefore largest in the high-$N_{\rm data}$ regime, where storage is most difficult. It combines many noisy measurements into a posterior mean visibility with an associated uncertainty. The result is a continuous, grid-free representation of the visibility that can be evaluated at arbitrary $(u,v)$ coordinates, rather than a compressed or averaged version of the original sampling.

\section*{CODE AVAILABILITY}
The code to reproduce the results presented in this paper is available at \url{https://github.com/takafumi291/uv_flow}. 

\acknowledgments
I acknowledge Jacki Ma, Hiep Nguyen, and Eric Muller for discussion over lunch at Baitong on 16 Apr 2024, during which I learned about the storage challenges facing the SKA and suggested the very idea developed in this work. I also thank Cameron Van Eck, Naomi McClure-Griffiths, and Satoru Iguchi for valuable discussions. This work was supported by Japan Foundation for Promotion of Astronomy and Center for Data-Driven Discovery (C3D) Seed Grant.

\bibliography{TsukuiSPIE26} 
\bibliographystyle{spiebib} 

\end{document}